\documentclass[12pt]{article}
\usepackage{amsmath}
\usepackage{graphicx}
\usepackage{natbib}
\usepackage{url} 

\newcommand{\blind}{1}

\usepackage{amssymb}
\usepackage{amsmath}
\usepackage{amsthm}
\usepackage{graphicx}
\usepackage{enumerate}

\newtheorem{theorem}{Theorem}[section]
\newtheorem{lemma}{Lemma}[section]

\newtheorem{assumption}{Assumption}[section]

\usepackage{mathtools}
\usepackage{float}
\usepackage{fancyhdr}

\usepackage{booktabs}
\usepackage{bm}
\usepackage{setspace}

\begin{document}

	\def\spacingset#1{\renewcommand{\baselinestretch}%
		{#1}\small\normalsize} \spacingset{1}

	
	\if1\blind
	{
		\title{\bf Diagnostics for Regression Models with Semicontinuous Outcomes}
		\author{Lu Yang
\\
			School of Statistics, University of Minnesota}
		\date{}
		\maketitle
	} \fi
	
	\if0\blind
	{
		\bigskip
		\bigskip
		\bigskip
		\begin{center}
			{\LARGE\bf Diagnostics for Regression Models with Semicontinuous Outcomes}
		\end{center}
		\medskip
	} \fi
	
	\bigskip

\begin{abstract}
	Semicontinuous outcomes commonly arise  in a wide variety of fields, such as  insurance claims, 
	healthcare expenditures, rainfall amounts, and alcohol consumption.
	Regression models, including Tobit, Tweedie, and two-part models, are widely employed to understand the relationship between semicontinuous outcomes and  covariates.   Given the potential detrimental  consequences of model misspecification,  after fitting a regression model, it is of prime importance to check the adequacy of the model.
	However, due to the point mass at zero, standard diagnostic tools for regression models (e.g., deviance and Pearson residuals)  are not informative for semicontinuous data. To bridge this gap, we propose a new type of residuals for semicontinuous outcomes  that are applicable to general regression models. Under the correctly specified model, the proposed residuals converge to being uniformly distributed, and when  the model is misspecified,  they significantly depart from this pattern.
	In addition to in-sample validation, the proposed methodology can also be employed to evaluate  predictive distributions.
	We demonstrate the effectiveness of the proposed tool using health expenditure  data from the  US Medical Expenditure Panel Survey.
	
\end{abstract}	

\noindent%
\vfill

\newpage
\spacingset{1.9} 

\section{Introduction}\label{sec:twintro}
Semicontinuous outcomes, also known as two-part outcomes, are commonly observed in various fields. For instance, in non-life insurance, claim outcomes generally follow a two-part distribution consisting of a probability mass at zero, indicating no claim, and the distribution of a positive-valued random variable describing the claim amount given occurrence \citep{frees2013actuarial,yang2020nonparametric}. Besides insurance claims,  examples of  semicontinuous outcomes   include healthcare expenditures with zero representing  no utilization  \citep{smith2017marginalized,huling2021two}, rainfall amounts with zero representing  no rain \citep{hyndman2000applications}, and alcohol consumption \citep{liu2008multi},     among many others.
There are other types of two-part data defined in the literature, such as zero-modified count data \citep{neelon2016modeling} and semicontinuous data with a continuous part that can   take both  positive and negative  values \citep{lu2004analyzing}.
In this paper, we focus on zero-inflated nonnegative continuous data due to their generality.

To understand the relationship between a semicontinuous outcome and a set of predictors,  two types of regression models are typically used (cf. \cite{xacur2015generalised,liu2019statistical}).
The first type is a two-part approach (\citealt{manning1981two,duan1983comparison}; \citealt*{frees2011predicting}), which models the zero and positive parts separately. 
This is also known as a hurdle or zero-inflated model  \citep{mullahy1998much}.
For example,  the probability mass at zero  can be characterized by a   logistic regression,
 and the positive part can be modeled using distributions of positive random variables such as lognormal, gamma, and generalized beta of the second kind (GB2) distributions
  \citep{mcdonald1995generalization}.
These two parts can be further linked using random effects \citep{olsen2001two,tooze2002analysis,farewell2017two}. Additionally, 
a double generalized linear model (GLM) can accommodate heteroscedasticity by modeling the dispersion using a linear combination of covariates  \citep{smyth1989generalized}.
 Marginalized two-part models \citep{smith2014marginalized}, in contrast,  characterize the covariate effects on the probability of zero and  the marginal mean.

The second type of methods is a 
one-part model. 
Tweedie distributions \citep{jorgensen1994fitting,ohlsson2006exact}, which assume that  the semicontinuous outcome is generated by a Poisson sum of gamma random variables, 
are widely employed to model semicontinuous claim outcomes in insurance. 
 Tweedie distributions belong to the exponential family, and thus the established results for GLMs can be  readily applied. 
The  mean and variance of a Tweedie variable satisfy the following relationship \begin{gather*}
	\mathrm{E}(Y)=\mu,~
	\mathrm{Var}(Y)= \phi \mu^{\pi},
\end{gather*}
where  $1<\pi<2$, and  $\phi$ is the dispersion parameter. 
	Under regression models, the mean $\mu$ is commonly  expressed  as a  function of linear combinations of  covariates, e.g., $\mu=\exp(\mathbf{X}^\top\bm\beta)$, where $\mathbf{X}$ is the vector of covariates and $\bm\beta$ is the vector of coefficients. 
Tobit models \citep{tobin1958estimation} assume that  the semicontinuous outcome is generated by a latent continuous variable through censoring and are  suitable when there exists a detection limit. 
Specifically, the zeros represent censoring of an underlying continuous variable $Y^*$ below a detection limit $L$. That is,
\begin{align*}Y=
	\begin{cases*}
 0& if $Y^* < L$ \\Y^* &if $Y^*\geq L$.
	\end{cases*}
\end{align*}
Customarily,  the latent variable is assumed to follow a normal distribution with mean  $\mathrm{E}(Y^*)=\mathbf{X}^\top \bm\beta.$
The  Heckman model \citep{heckman1976common}, or Type II Tobit 
model,   closely associates with the two-part model, further allowing the probability of  censoring and the observed response variable to
relate to different covariates and errors.

After fitting a regression model,
model assessment and diagnostics  are  routine statistical tasks, in particular in highly regulated fields such as insurance and healthcare. 
Model misspecification might induce numerous potential
detrimental  consequences. 
For instance,  in insurance practice, an inappropriate risk model can lead to catastrophic
insolvency. Therefore, analysts are 
often  tasked with evaluating the model, including  inspecting whether the distribution family is appropriate and determining whether there are other important  predictors which should be considered.

Residuals  play a crucial role in regression model diagnostics. 
Pearson and deviance residuals are  widely  employed  for this purpose \citep{mccullagh1989generalized}. 
In  linear regression models with normal errors, if the model is correctly specified, Pearson and deviance residuals should closely follow a normal distribution. 
One can evaluate the adequacy of the fitted model by comparing the distribution of residuals with the  hypothesized pattern (i.e., a normal distribution). A significant discrepancy indicates a deficiency in the fitted model. 
 Beyond normality, Cox-Snell residuals \citep{cox1968general} are suitable for regression models with continuous outcomes  (e.g., gamma).
 For  continuous outcomes  $Y_i,~i=1\ldots,n$, Cox-Snell residuals are defined as 
$\hat{F}_{i}(Y_{i}), i=1,\ldots,n$, where  $\hat{F}_i$ is the  fitted cumulative distribution function of $Y_i$.
If the model is correctly specified,  Cox-Snell residuals should exhibit a uniform trend.

\begin{figure}[!htbp]
	\centering
\includegraphics[width=\textwidth]{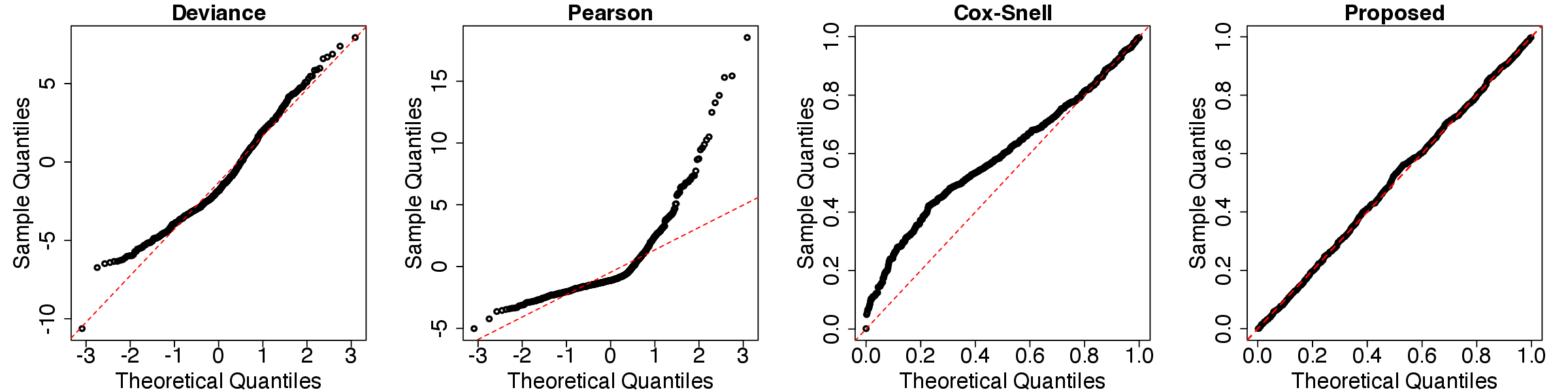} 

	\caption{QQ plots of deviance, Pearson,  Cox-Snell, and the propose residuals for a simulated Tweedie  example when the model is correctly specified. \label{fig:residuals}}
\end{figure}

However, the standard residuals 
may  not  be informative for semicontinuous outcomes.    For demonstration, Figure \ref{fig:residuals} includes the QQ plots of deviance,  Pearson, and Cox-Snell residuals for simulated  Tweedie data 
when the regression model is correctly specified.
Strikingly,   the residuals show significant deviations from their expected patterns (normality for deviance and Pearson residuals and uniformity for Cox-Snell residuals), even when the model is correctly specified.  This indicates that observing large discrepancies between these residuals and their hypothesized patterns does not necessarily imply  model deficiency,
rendering these conventional residuals unreliable diagnostic tools for semicontinuous data.

Despite of the prevalence of semicontinuous data,
diagnostics for regression models with semicontinuous outcomes have not been systematically studied in the literature.
One naive approach is to assess the model fitting of  zero and positive parts separately \citep{shi2018pair}. However, this method is   not ideal  because, firstly, it is not sensible  to separate these  parts if one adopts a one-part model such as a Tweedie GLM, since one cannot adjust the fitting of each component individually. Secondly,  diagnostics for regression models with binary outcomes are particularly known to have many open issues \citep{yang2021assessment}.
 Another ad hoc approach to tackle semicontinuous data 
is jittering,
which  transforms the semicontinuous outcomes into continuous outcomes by  adding a simulated uniform random variable.    Subsequently, Cox-Snell residuals of the transformed outcomes can be obtained   \citep{dunn1996randomized,yang2019multiperil}. 
However, this procedure injects noise into the data, making it  contingent  on the realization of the noise. 
As a result, even with the same data and model, different  realizations of the random noise can produce different sets of residuals. We illustrate this through the simulation study in the supplementary materials. 
%
%
%

    In current practice, 
analysts often use information criteria such as AIC and BIC or techniques such as cross validation for model selection.  
However, these tools do not provide statistical statements on model adequacy. That is,
even  the best model among candidates is not guaranteed to fit the data sufficiently. Hence,  assessing the adequacy  of the selected model may be a further task after completing  model comparison.
  While  Lorenz curves and  Gini indices have been used to  evaluate the mean structure of insurance claims   \citep*{frees2011summarizing}, our goal is to assess the entire distribution beyond the first moment.

In this article, we propose a new type of residuals specifically tailored for regression models with semicontinuous outcomes. 
 When the model is correctly specified, the proposed  residuals closely follow a uniform distribution.  
In cases of  misspecification, such as underestimating tail probabilities and missing covariates, the proposed residuals deviate significantly from the expected pattern, indicating a need for model refinement.  The shape of  residuals plots can further hint at potential causes of misspecification.
Our residuals are broadly applicable to  regression models for semicontinuous data, including both one-part and two-part models.

The remainder of the article is organized as follows. In Section \ref{sec:twuni}, we propose new residuals for semicontinuous outcomes. The theoretical properties of the proposed residuals are established in Section \ref{sec:theo}.  In Section \ref{sec:twsimu}, we  evaluate the performance of the proposed residuals under different scenarios by means of a simulation study, and in Section \ref{twdata}, we demonstrate the usage of our tool on a real dataset from the   US Medical Expenditure Panel Survey. The article closes with a discussion in Section \ref{conclusion}. Technical details are presented in the appendix.

\section{New Residuals for Regression Models with Semicontinuous Outcomes}\label{sec:twuni}

Let $Y$ follow a semicontinuous distribution with probability $p_0$ of being zero
 and density function $g$ for its positive part.
In a regression model, $p_0$ and $g$ are associated with covariates $\mathbf{X}$. 
In  one-part models, one set of covariates is used to fit $p_0$ and $g$  simultaneously, whereas they may have different covariates and corresponding coefficients  in a two-part model.

Let $\delta_0$ be the Dirac measure at 0, 
	and $m$ be the Lebesgue measure.
Then given $\mathbf{X}=\mathbf{x}$, the conditional density of $Y$  with respect to $m+\delta_0$ is $$f(y|\mathbf{x})=\begin{cases*}
p_0(\mathbf{x})&$y=0$\\\left(1-p_0(\mathbf{x})\right)g(y|\mathbf{x})&$y>0$,
\end{cases*}$$
and its cumulative distribution function is 
\begin{align}\label{eq:cdf}
F(y|\mathbf{x})=\Pr(Y\leq y|\mathbf{X}=\mathbf{x})=p_0(\mathbf{x})+\left(1-p_0(\mathbf{x})\right)G(y|\mathbf{x})
\end{align}
where $G$ is the cumulative distribution function associated with $g$.

For a continuous variable $Y$ whose distribution function is $F$, the probability integral transform $F(Y|\mathbf{X})$ is uniformly distributed. 
This forms the basis for Cox-Snell residuals,  $\hat{F}(Y_i|\mathbf{X}_i),i=1,\ldots,n,$
where  $\hat{F}$ is the  fitted distribution function. 
The degree of proximity between  Cox-Snell residuals and the uniform distribution reflects the adequacy of the fitted model for continuous outcomes.
However, as shown in the third panel of Figure \ref{fig:residuals}, Cox-Snell residuals are not uniformly distributed for semicontinuous  data even if the model is correctly specified.

Our goal is to  develop a diagnostic tool for semicontinuous data.
To gain insights, we  analyze the probability integral transform $F(Y|\mathbf{X})$ of a semicontinuous variable. We first condition on $\mathbf{X}=\mathbf{x}$ and obtain $\Pr\left(F(Y|\mathbf{X})\leq s|\mathbf{X}=\mathbf{x} \right)$, for a fixed point $s\in (0,1)$. Then we take the expectation with respect to $\mathbf{X}$ to acquire $\Pr\left(F(Y|\mathbf{X})\leq s\right)$.

Since $F(Y|\mathbf{x})=p_0(\mathbf{x})+\left(1-p_0(\mathbf{x})\right)G(Y|\mathbf{x})\geq p_{0}(\mathbf{x})$,
if $p_0(\mathbf{x})>s$, then $F(Y|\mathbf{x})\geq p_0(\mathbf{x})>s$, and thus $\Pr\left(F(Y|\mathbf{x})\leq s \right)=0$. On the other hand, if $p_{0}(\mathbf{x})\leq  s$, 
the event that $F(Y|\mathbf{x})=p_0(\mathbf{x})+\left(1-p_0(\mathbf{x})\right)G(Y|\mathbf{x})\leq s $ is equivalent to 
$Y\leq G^{-1}\left(\left.\frac{s-p_{0}(\mathbf{x})}{1-p_{0}(\mathbf{x})}\right|\mathbf{x}\right).$ Applying \eqref{eq:cdf}, the corresponding  probability is 
\begin{align*}
\Pr\left(F(Y|\mathbf{x})\leq s|\mathbf{X}=\mathbf{x}\right)=
&\Pr\left(\left.Y\leq G^{-1}\left(\left.\frac{s-p_{0}(\mathbf{x})}{1-p_{0}(\mathbf{x})}\right|\mathbf{x}\right)\right|\mathbf{X}=\mathbf{x} \right)\\=&p_{0}(\mathbf{x})+\left(1-p_{0}(\mathbf{x})\right)G\left(\left.G^{-1}\left(\left.\frac{s-p_{0}(\mathbf{x})}{1-p_{0}(\mathbf{x})}\right|\mathbf{x}\right)\right|\mathbf{x}\right)\\=&s, \text{ for }p_{0}(\mathbf{x})\leq  s.
\end{align*}
Combination  yields that
\begin{align*}
\begin{split}
\Pr\left(F(Y|\mathbf{X})\leq s |\mathbf{X}=\mathbf{x}\right)
&=\begin{cases*}
0&$p_{0}(\mathbf{x})>s$,\\
s&$p_{0}(\mathbf{x})\leq  s$.
\end{cases*}
\end{split}
\end{align*}
Take the expectation with respect to $\mathbf{X}$ to get 
\begin{align}\label{eq:hfunc}
H(s)\coloneqq\Pr(F(Y|\mathbf{X})\leq s)=
0\cdot\Pr\left(p_{0}(\mathbf{X})>s\right)+s\Pr\left(p_{0}(\mathbf{X})\leq s\right)
=s\Pr(p_0(\mathbf{X})\leq s).
\end{align}
For semicontinuous outcomes, $s\Pr(p_0(\mathbf{X})\leq s)\neq s$.
This elucidates that the probability integral transform  of a semicontinuous outcome does  not follow a  uniform distribution  due to the presence of a point mass at zero. 
This is reflected in the third panel of Figure \ref{fig:residuals}.

If $\mathbf{X}$ contains  continuous components, $F(Y|\mathbf{X})$ is a continuous variable whose   cumulative distribution function is $s\Pr(p_0(\mathbf{X})\leq s)$. Therefore, although $F(Y|\mathbf{X})$ is not  uniformly distributed, another layer of probability integral transform, namely $H\left(F(Y|\mathbf{X})\right)$, yields a uniform variable. We construct our residuals based on the empirical counterpart of $H\left(F(Y|\mathbf{X})\right)$.

Now suppose we have a sample $(\mathbf{X}_i,Y_i),i=1,\ldots,n, $ as independent realizations of $(\mathbf{X},Y)$.
We  first estimate the regression model using all the data to obtain $\hat{p}_0$ and $\hat{F}$. 
Then we estimate the function $H$ defined in \eqref{eq:hfunc} using its empirical estimator \begin{align}\label{eq:h}
	\hat{H}(s)=\frac{s}{n}\sum_{j=1}^n1\left(\hat{p}_0(\mathbf{X}_j)\leq s\right).
\end{align}
Combining,
we  propose    residuals 
\begin{align}\label{eq:propose}
\hat{r}_i=\hat{H}\left(\hat{F}(Y_i|\mathbf{X}_i)\right)=\frac{\hat{F}(Y_i|\mathbf{X}_i)}{n}\sum_{j=1}^n1\left(\hat{p}_0(\mathbf{X}_j)\leq \hat{F}(Y_i|\mathbf{X}_i)\right), i=1,\ldots,n.
\end{align}
The proposed residuals $\hat{r}_i$ resembles $H\left(F(Y|\mathbf{X})\right)$, which has a null distribution of uniformity. Therefore, the   degree of agreement between the proposed residuals and uniformity reflects the goodness-of-fit of the model.
Importantly, our residuals only require $\hat{p}_0(\mathbf{X}_i)$ and $\hat{F}(Y_i|\mathbf{X}_i)$ as input, making them 
  applicable to general regression  models for semicontinuous outcomes.


Furthermore, one can conduct a  normal quantile transformation on the residuals. That is, $\Phi^{-1}(\hat{r}_i),~i=1,\ldots,n$, where $\Phi^{-1}$ is the quantile function of a standard normal variable. By doing so, the null pattern becomes the standard normal distribution. 
The resulting visualization is useful 
 for examining the tail behavior of the residuals.

For quick demonstration, 
we continue with the example presented in Figure \ref{fig:residuals} and show the proposed residuals the right panel.  As can be seen,  the proposed residuals closely follow the null pattern under the correctly specified model. This indicates that our proposed residuals can serve as a reliable tool for diagnosing regression models with semicontinuous data.  A thorough simulation study is included in Section \ref{sec:twsimu}.

%

%
%
%

%
%

\subsection{Theoretical Properties}\label{sec:theo}

Now we establish the theoretical properties of   the proposed residuals. 
For clarity, we index  functions with their parameters ${\bm\beta}$ in this section. 
When the coefficients are set to be $\bm\beta$, we denote the resulting  distribution function in \eqref{eq:cdf} as $F(\cdot| \mathbf{X},\bm\beta)$ and the probability of zero as $p_0(\mathbf{X},\bm\beta)$. 
The resulting $H$ in \eqref{eq:hfunc} and  its estimator $\hat{H}$ in \eqref{eq:h} and are denoted as $H_{{\bm\beta}}$ and $\hat{H}_{{\bm\beta}}$, respectively.
The underlying parameters ${\bm\beta}_0$ are estimated using the maximum likelihood estimator $\hat{  \bm\beta}$.
%

We focus on the null behaviors of the proposed residuals,
namely when the model is correctly specified and thus  $Y|\mathbf{X}=\mathbf{x}\sim F(y|\mathbf{x},\bm{\beta}_0)$.
As discussed in Section \ref{sec:twuni},  $H_{{\bm\beta}_0}\left(F(Y|\mathbf{X},{\bm\beta}_0)\right)$, which can be viewed as the  model error, follows a uniform distribution under the true model.
We are interested in the difference between the residual and the  model error $${\sqrt{n}}\left[\hat{H}_{\hat{  \bm\beta}}\left(F(Y|\mathbf{X},\hat{{\bm\beta}})\right)-H_{{\bm\beta}_0}\left(F(Y|\mathbf{X},{\bm\beta}_0)\right)\right].$$

\begin{theorem}\label{theorem}
	Under Assumptions \ref{op}, \ref{assume:margin}, \ref{lips}, and \ref{monof} given in the appendix, when the model is correctly specified, the  limiting  distribution  of  ${\sqrt{n}}\left[\hat{H}_{\hat{  \bm\beta}}\left(F(Y|\mathbf{X},\hat{{\bm\beta}})\right)-H_{{\bm\beta}_0}\left(F(Y|\mathbf{X},{\bm\beta}_0)\right)\right]$ has the representation   $\mathbb{G}h_{\mathbf{X},Y}$, where $\mathbb{G}$ is a  Brownian bridge, and
	$$h_{\mathbf{X},Y}(\mathbf{x}_i,y_i)=
	F(Y|\mathbf{X},{\bm\beta}_0)1\left(p_0(\mathbf{x}_i,{\bm\beta})\leq F(Y|\mathbf{X},{\bm\beta}_0)\right)
	+\left.\frac{\partial H_{{\bm\beta}} \left( F(Y|\mathbf{X},{\bm\beta})\right)}{\partial \bm\beta}\right\vert_{\bm\beta=\bm\beta_0}\left[I(\bm\beta_0)\right]^{-1}\dot{l}_{\bm\beta_{0}}(\mathbf{x}_i,y_i).$$
	
\end{theorem}

Theorem \ref{theorem} guarantees that 
the proposed residuals asymptotically converge to the model error 
 with the order  ${n}^{-1/2}$.
This implies that when the model is correctly specified, the proposed residuals closely resemble  a uniform distribution, 
providing theoretical justification for  their use as a diagnostic tool for regression models with semicontinuous outcomes.
  As shown in  Theorem \ref{theorem}, the discrepancy between the residuals and the errors arises from estimation, with the first part resulting from the estimation of $H$ and the second part induced by the estimation of $\bm{\beta}_0$.
The proof of the theorem is provided in the appendix.


\subsection{Evaluation of Predictive Distributions}
Our framework  can  be extended for out-of-sample validation. Suppose we have fit  $\hat{F}$ and  $\hat{p}_{0}$ using the in-sample data, 
 and $(\tilde{\mathbf{X}}_i,\tilde{Y}_i),i=1,\ldots,m$, are the of out-of-sample data. 
 For the $i$th out-of-sample observation, $\hat{F}\left(y|\tilde{\mathbf{X}}_i\right)$ is the predictive distribution function, and  $\hat{p}_0(\tilde{\mathbf{X}}_i)$ is the predictive probability of zero.
 To  evaluate whether  $\hat{F}$ and $\hat{p}$ describe the distribution of the out-of-sample data,
 one can obtain the out-of-sample errors
 $$\tilde{r}_i=\frac{\hat{F}\left(\tilde{Y}_i|\tilde{\mathbf{X}}_i\right)}{m}\sum_{j=1}^m1\left(\hat{p}_0(\tilde{\mathbf{X}}_j)\leq \hat{F}\left(\tilde{Y}_i|\tilde{\mathbf{X}}_i\right)\right),i=1,\dots,m.$$
 If  the
 held-out sample indeed follows the predictive distribution, the out-of-sample errors are expected to be close to being uniformly distributed.

In the literature, 
there are tools available for assessing point predictions, 
for instance, 
the  mean absolute prediction error,  
the  mean squared prediction error, and 
the Gini index 
\citep*{frees2011summarizing}. 
Our proposed tool goes beyond point predictions and assesses the adequacy of the entire predictive distribution at the individual level. 

\section{Simulation}\label{sec:twsimu}

In the simulation study, we investigate the finite-sample performance of the proposed residuals. Specifically, we examine their proximity to the null pattern  under the correct model  and sensitivity to misspecification, which forms the foundation for an effective diagnostic tool.
In Section \ref{sec:dismiss}, we investigate misspecification in the distribution family, while in Section \ref{sec:misscov}, we explore the issue of missing covariates. Additionally, we analyze the impacts of sample sizes,  proportions of zeros,   correlations among covariates, and other factors that might affect the behavior of the proposed residuals.

\subsection{Incorrect Distribution Function}\label{sec:dismiss}

We first consider a situation where the distribution family is misspecified,  
demonstrating the usage of our tool in choosing between one-part and two-part models.
The underlying model  is a two-part model. The probability of zero is  $$p_0(\mathbf{X})=\text{logit}^{-1}\left(\beta_0+X_{1}\beta_{1}+X_{2}\beta_{2} \right),$$
where $X_1$ is a standard normal variable,  $X_2$ is binary with probability of one as 0.4, and $(\beta_{1},\beta_{2})=(-2,-1)$.  In the following sections, we vary the value of $\beta_{0}$ to adjust the proportion of zeros.
A gamma distribution is employed to generate positive data. The mean function of the positive part is described as $$\lambda_S=\exp\left(\beta_{0S}+\beta_{1S}X_1+\beta_{2S}X_2\right).$$ We let  $(\beta_{0S},\beta_{1S},\beta_{2S})=(-1,-1,-2)$. The dispersion parameter is set to be 0.5. In the supplementary materials, we further explore a long-tailed distribution example in which a GB2 distribution is employed to generate  positive data.
 To facilitate visualization, we apply the quantile function of the standard normal distribution  to the proposed residuals in most examples. Consequently,  a standard normal distribution serves as the null pattern.

\subsubsection{Effect of Sample Size}
\begin{figure}[!htbp]
	\centering
	\includegraphics[width=.9\textwidth]{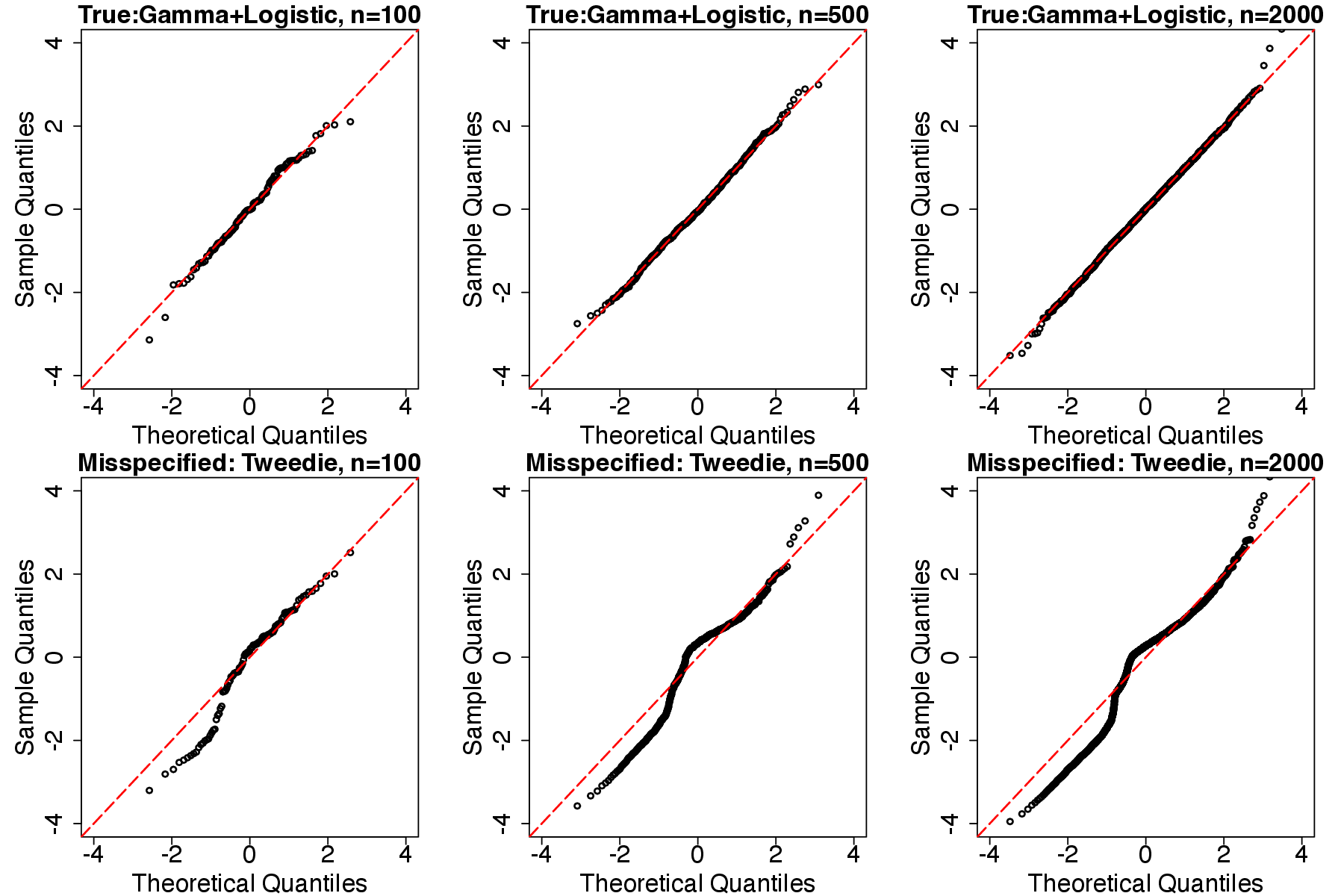} 
	
	\caption{QQ plots of the proposed residuals under the true model, gamma+logistic, in the top row compared with the misspecified model,  a Tweedie model,  in the bottom row. 
		The sample sizes are 100, 500, and 2000 in the three columns.
		\label{fig:gammasize}}
\end{figure}

We first let $\beta_0=-1$, resulting in approximately $30\%$ of the data being zeros. 
We consider three sample sizes, 100, 500, and 2000, to  explore their effects. Figure \ref{fig:gammasize} presents the results. 
Our residuals are  very close to being normally distributed under the true model in the top row and deviate significantly  from the null pattern  under the misspecified model in the bottom row. This is true even for a  small sample size of 100 in the left column.
As the sample size increases from 100 to 500 and 2000, we observe an evident improvement in the performance of the proposed residuals.  They exhibit a stronger resemblance to the diagonal under the true model and  a more pronounced deviation under misspecification. 

\subsubsection{Effect of Zero Proportions}
\begin{figure}[!htbp]
	\centering
	\includegraphics[width=.9\textwidth]{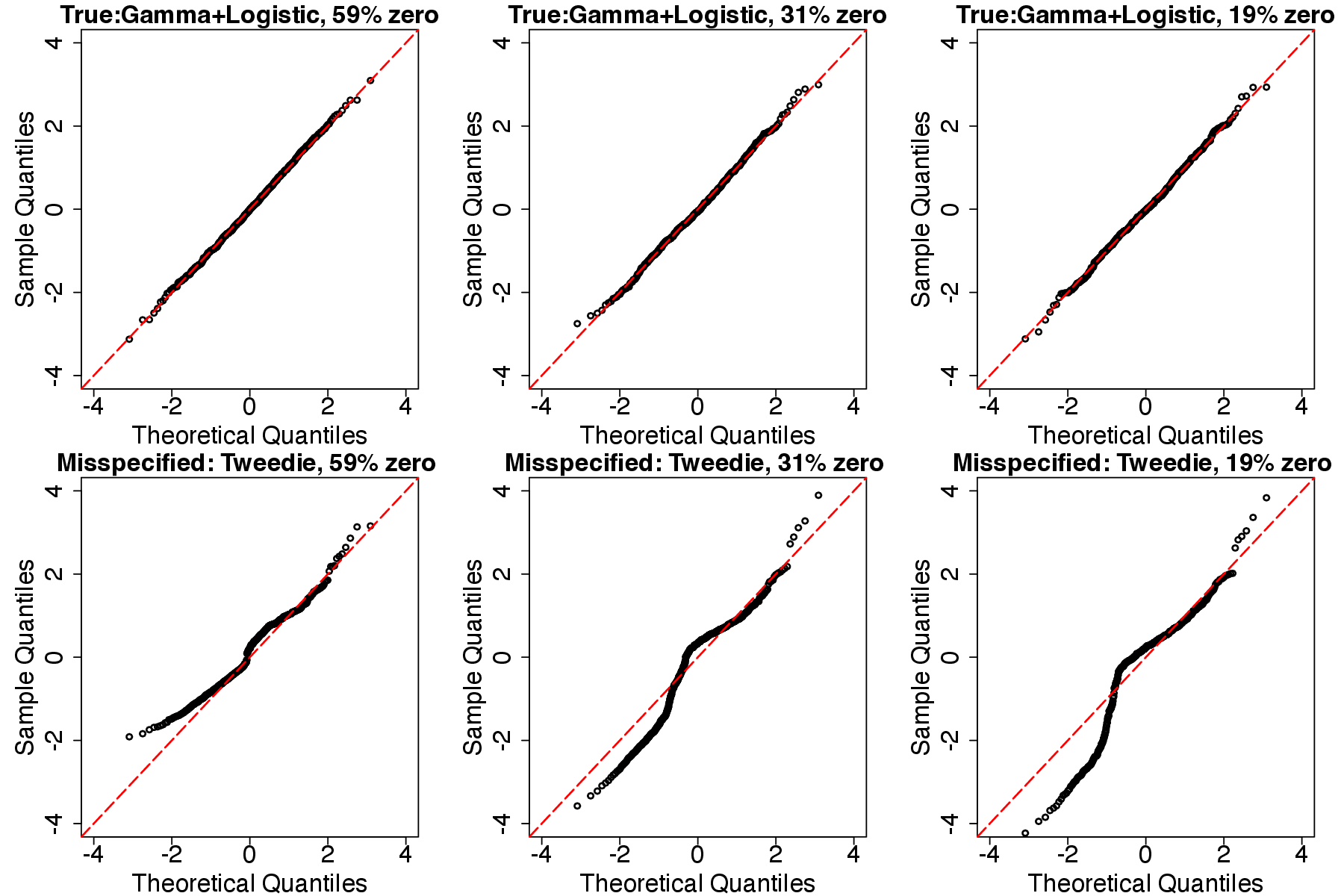} 
	
	\caption{QQ plots of the proposed residuals under the true model, gamma+logistic, in the top row compared with the misspecified model,  a Tweedie model,  in the bottom row. 
		The zero proportions are 59\%, 31\%, and 19\% in the three columns. The sample size is 500.
		\label{fig:gammazero}}
\end{figure}

We further investigate the effect of zero proportions by adjusting the value of $\beta_{0}$ to be 1, $-1$, and $-2$. Correspondingly, the zero proportions are about 59\%, 31\%, and 19\%. We can see from Figure \ref{fig:gammazero} that the proportion of zeros impacts the pattern of the residuals under the misspecified model. 
The deviation with the diagonal is more apparent with fewer zeros (right column) compared to scenarios with  many zero  (left column).
When there are many zeros, the zero part is dominant in the behavior of the residuals. However, 
  diagnostics for regression models with binary outcomes are known to be challenging due to the limited information available in the outcomes \citep{yang2021assessment}.

\subsection{Missing Covariates}\label{sec:misscov}
In this section, 
we explore the application of the proposed residuals in  Tobit models outlined in Section \ref{sec:twintro}.
We  investigate the ability of the proposed residuals to detect missing covariates in various scenarios. 

\begin{figure}[!htbp]
	\centering
	\includegraphics[width=.9\textwidth]{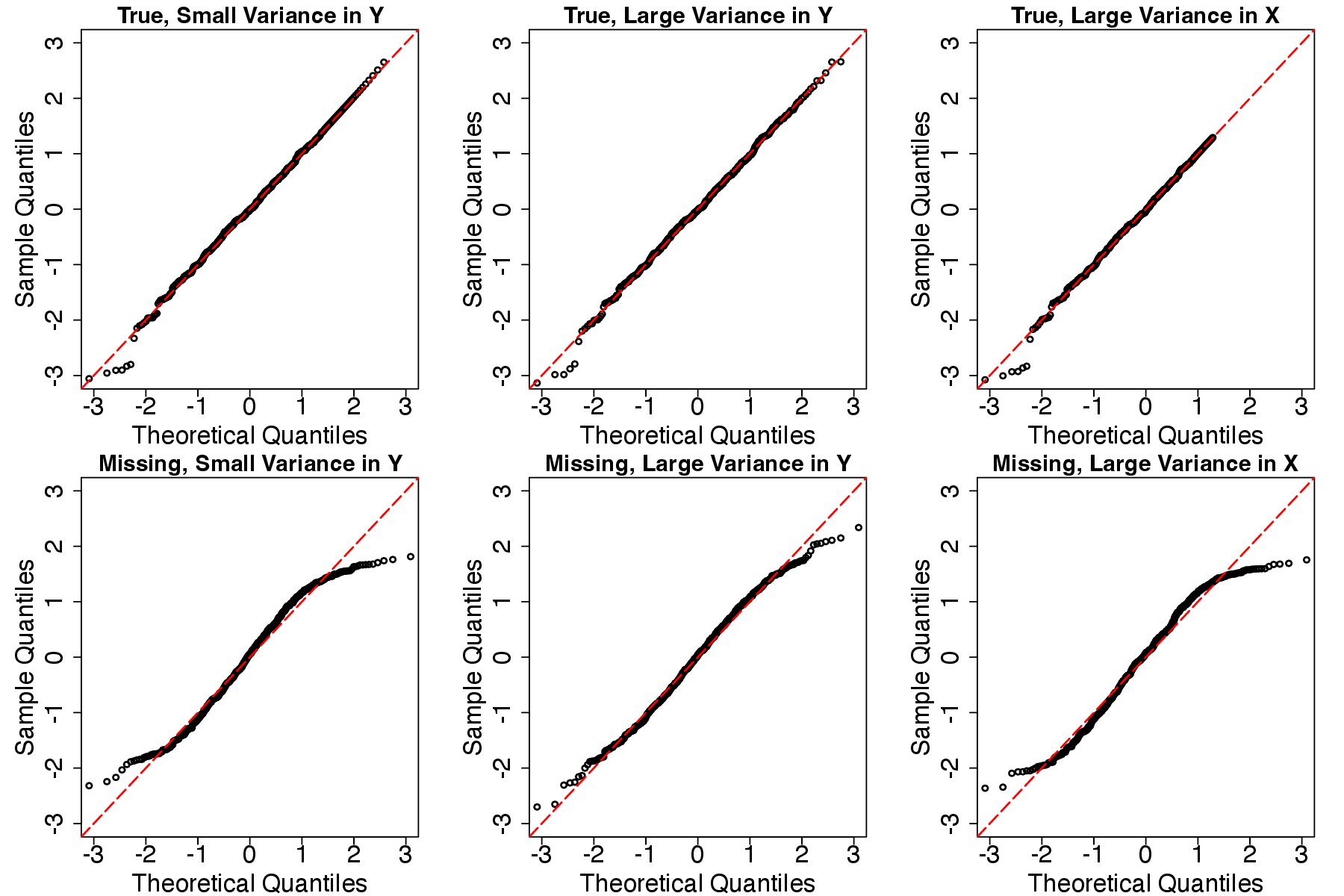} 
	
	\caption{QQ plots of the proposed residuals for Tobit models. The model is correctly specified in the top row, and in the bottom row, one covariate is missing.
		 The sample size is  500.\label{fig:missingmain}}
\end{figure}

We assume that the latent variable  $Y^*$ follows a normal distribution with a mean given by
$$\mu=\beta_0+\beta_1X_1+\beta_2X_2, $$ where $X_1, X_2 \sim Unif (-1, 1)$ independently, and $\beta_0 = 2, \beta_1 = 2, \beta_2 = 2$. 
We observe $Y=0$ if $Y^*<0$, and around 15\% of the outcomes are zeros.
Under the misspecified model, the covariate $X_2$ is missing. 
We vary the standard deviation of the latent variable to be 0.2 (left column of Figure \ref{fig:missingmain}) and 0.6 (middle column) to explore its effect.
In addition, in the right panel of Figure \ref{fig:missingmain}, we explore the effects of the variance in the covariates by letting $X_1, X_2 \sim Unif (-1.5, 1.5)$ instead.

From Figure \ref{fig:missingmain}, we  can see that the proposed residuals closely follow their null pattern when the model is correctly specified in all the scenarios. A low noise level, corresponding to a small variance in the latent variable, tends to give a strong signal of model misspecification (left column). In contrast, the misspecification is masked when the noise is more dominant in the middle column.
On the other hand, the performance of the proposed residuals is not significantly impacted by the distribution of covariates,  as evident in the comparison between the left and right columns. This is attributed to the fact that we assess the conditional distribution of $Y|\mathbf{X}.$

In the supplementary materials, we illustrate that our residuals can  detect missing interactions. Furthermore, they prove effective  even in the presence of mild or moderate dependence among the covariates.

\section{Medical Expenditure Data}\label{twdata}
In this section, 
 we analyze  data from the  US Medical Expenditure Panel Survey  (MEPS),
 a set of extensive surveys of   individuals,  medical providers, and employers across the United States. The MEPS includes complete data on  healthcare service utilization,  
as well as    information on respondents' health status, demographic and socioeconomic characteristics, employment, access to  care, among other factors.

We focus on  medical costs related to 
office-based (OB) visits, which is one of the most  frequently utilized healthcare services.
The  outcome of interest is the aggregate  annual OB expenditure per  participant, including out-of-pocket payments by users and payments made by other sources  such as insurance companies. 
Medical expenditures  naturally come in two parts, with zeros  arising from individuals without medical utilization. 
We analyze the expenditure data from 2018 and reserve the data from 2019 for out-of-sample validation. 

\subsection{Data Summary}
Table \ref{tab:dist} summarizes the distribution of the OB expenditures in 2018  for a total of  29,784 participants.  Around  $65.8\%$ of the participants did not incur any medical costs in 2018. In addition, we  provide the quantiles of the cost distributions given utilization, which reveal the right skewness and long tails of the expenditure distribution.

\begin{table}[!htbp] \centering 
	\caption{Zero proportion and percentiles of positive expenditures (in dollars) for OB services in 2018.\label{tab:dist} }
	\begin{tabular}{@{\extracolsep{5pt}} cccccccc} 
		\\[-1.8ex]\hline 
		\hline \\[-1.8ex] 
		Zero proportion&min& 25\% & 50\% & 75\% & 90\% & 95\% & max\\ 
		\hline \\[-1.8ex] 
		$0.658$ &   2   & 189&    455  & 1,184   &2,929   &5,094& 268,552  \\ 
		\hline \\[-1.8ex] 
	\end{tabular} 
\end{table}

A table  of potential covariates and their  summary statistics is included in the supplementary materials.
Following \cite{frees2013actuarial},
we consider various demographic characteristics including age, gender, and ethnicity as well as region   variables which provide  insight into  overall economic and environmental conditions. We also consider  socioeconomic factors  including education, marital status, family size,  income, employment,  and industry. Health status variables including physical and mental health indicators and activity limitation are also included as      relevant factors. Lastly, enrollment in insurance and managed care plan can  impact  the out-of-pocket payment and thus is expected to have an important implication on health expenditures.

\subsection{Regression Models}\label{sec:reg}
To accommodate the two-part feature present in the OB expenditures, 
we fit the data with regression models described in Section \ref{sec:twintro}.  Specifically, we consider   a Tweedie GLM as well as  two-part models that combine logistic regression with either a gamma GLM or a GB2 distribution.

The coefficients of the fitted models are included in 
the supplementary materials.
The results  summarized in these tables suggest that   the  covariates indeed have different effects on the zero and positive parts,
as discussed in \cite*{frees2011predicting} and \cite{frees2013actuarial}. 
One can view a Tweedie model as a compromise between the zero and positive parts \citep{xacur2015generalised}. 

\subsection{Model Diagnostics}
\begin{figure}[!htbp]
	\centering
	\includegraphics[width=.6\textwidth]{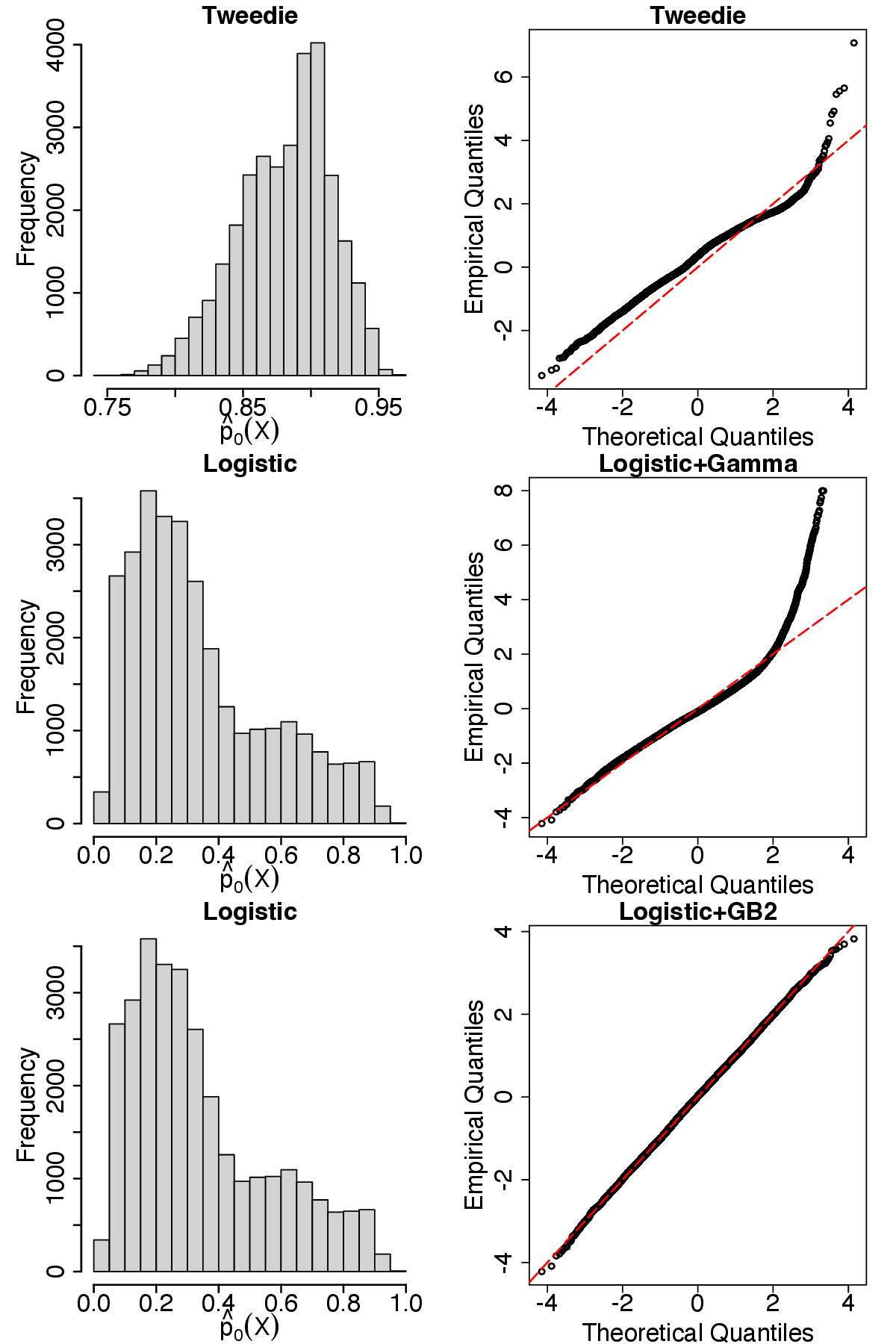} 
	\caption{Histograms of fitted $\hat{p}_0(\mathbf{X})$ (left columns) and QQ plots of the residuals (right column) from the Tweedie GLM (top row), two-part model with a gamma distribution (middle row), and two-part model with  a GB2 distribution (bottom row) on the OB expenditures. \label{fig:datamargin}}
\end{figure}

Given the fitted regression models,  we now apply the proposed residuals in Section \ref{sec:twuni} for model diagnostics.  
Figure \ref{fig:datamargin} presents the histograms of the fitted probabilities of zero $\hat{p}_0(\mathbf{X})$ (left column) and the  QQ plots of the proposed residuals (right column) for each model. 
The top row of Figure \ref{fig:datamargin} illustrates the poor performance of the Tweedie model. According to the Tweedie model, the fitted probabilities of no visit are mostly distributed above 0.7, 
 which contradicts Table \ref{tab:dist} that shows  the overall zero proportion
to be 0.658. As discussed earlier, the result of the Tweedie distribution can be viewed as a compromise between the zero and positive parts. 
The distribution of the zero part is perhaps distorted by the positive part.
Additionally,   the 
upper tail of the residuals  deviates from the diagonal,
indicating that the positive part is not fit well either.

The middle and bottom rows  of Figure \ref{fig:datamargin} display the  residuals of the two-part models. The combination of a logistic regression and the GB2 distribution seems to outperform the other models, with  
residuals closely following the null pattern.
In contrast, the right tail of the gamma regression residuals shows a large discrepancy with the diagonal, indicating that the positive expenditures have a longer tail than what is assumed by a gamma distribution.

%
\subsection{Out-of-Sample}

We can obtain the predictive distribution of each individual's expenditure in 2019 using the fitted models in Section \ref{sec:reg} together with the available covariates in the held-out sample of 2019. 
We present the histograms of the predictive probabilities of zero and  the QQ plots of the out-of-sample  errors in the supplementary materials. 
Consistent with our in-sample analysis, we can conclude that the logistic regression combined with the {GB2} distribution outperforms other models. Although the GB2 distribution is more sophisticated than the gamma distribution with more parameters, overfitting  is not a concern in our application.  In contrast, the Tweedie model provides a poor out-of-sample fit.

\section{Conclusions}\label{conclusion}
In this article, we  introduced a new type of residuals for  assessing regression models with semicontinuous outcomes.
The proposed residuals converge to being uniformly distributed  when the model is correctly specified. Through simulations, we showed that the proposed residuals form an effective diagnostic tool, since they are close to the null pattern (a uniform distribution) under the correctly specified model, and  deviate significantly from this pattern when the model is misspecified. We illustrated that the proposed tool can effectively detect common causes of misspecification, including missing covariates and incorrect distributions. We also explored the effects of sample sizes, proportions of zeros, correlations in covariates, and other factors on the behavior of the residuals.

We conducted a case study on the MEPS data. Using our  tool, we concluded  that the combination of  logistic and GB2  regressions provides a good fit for our data, while the Tweedie GLM fits the data insufficiently. In addition to in-sample analysis, we also demonstrated that the proposed tool can be adopted to evaluate the  individual level predictive distribution. 

\section*{Acknowledgments}
The author gratefully acknowledges support from NSF grant DMS-2210712.

\section*{Supplementary Materials}

Web Figures and Tables referenced in Sections \ref{sec:twsimu} and \ref{twdata} 
as well as data and code needed to reproduce the results are available with this paper at the Biometrics website on Oxford Academic.  

\section*{Data Availability}
The Medical Expenditure Panel Survey data analyzed in this paper are openly available from the Agency for Healthcare Research and Quality at \texttt{https://meps.ahrq.gov/mepsweb/}.
\bibliographystyle{biom} 

\bibliography{nontweedie}

\appendix


\section{Proofs}\label{sec:proof}
\begin{assumption}\label{op}
	
	$$n^{1/2}(\hat{ \bm \beta}-  \bm\beta_0)\rightarrow N(0,[I(\bm\beta_0)]^{-1}),$$ where $I(\bm\beta)$ is the Fisher information matrix $P\left(\dot{l}_{\bm\beta}\dot{l}_{\bm\beta}'\right)$. Moreover, ${l}_{\bm\beta}(\mathbf{x},y)$ is   the log-likelihood  of the regression models, and $\dot{l}_{\bm\beta}(\mathbf{x},y)=\partial  l_{\bm\beta}(\mathbf{x},y)/\partial\bm\beta$ is the score function. 
\end{assumption}
\begin{assumption}\label{assume:margin}
	$p_0(\mathbf{X},\bm\beta)$, whose distribution depends on the distribution of  $\mathbf{X}$, has bounded density.
\end{assumption}

\begin{assumption}[Lipschitz condition]\label{lips}There exists a constant $\alpha_1$ such that for  $\bm\beta,\bm\beta'\in B$,  
	\begin{gather*}
		\left|p_0(\mathbf{x},\bm\beta)-p_0(\mathbf{x},\bm\beta')\right|\leq  \alpha_1\left|\bm\beta-\bm\beta'\right|,
		\\
		\left|F(y|\mathbf{x},\bm\beta)-F(y|\mathbf{x},\bm\beta')\right|\leq  \alpha_1\left|\bm\beta-\bm\beta'\right|,
	\end{gather*}
	where $B$ is the Euclidean parameter space.
\end{assumption}

\begin{assumption}\label{monof} 
	
	
	$H_{{\bm\beta}}\left(F(y|\mathbf{x};{\bm\beta})\right)$ is differentiable with respect to $\bm\beta$ for $\bm\beta\in B$,  and the derivatives are bounded. 

\end{assumption}


We decompose the difference between the residual and error into
\begin{align}\label{eq:hn}
	\begin{split}
		&{\sqrt{n}}\left[\hat{H}_{\hat{  \bm\beta}}\left(F(y|\mathbf{x},\hat{{\bm\beta}})\right)-H_{{\bm\beta}_0}\left(F(y|\mathbf{x},{\bm\beta}_0)\right)\right]\\
		=&{\sqrt{n}}\left[\hat{H}_{{  \bm\beta}_0}\left(F(y|\mathbf{x},{{\bm\beta}_0})\right)-H_{{\bm\beta}_0}\left(F(y|\mathbf{x},{\bm\beta}_0)\right)\right]\\&+{\sqrt{n}}\left[\hat{H}_{\hat{{\bm\beta}}}\left(F(y|\mathbf{x},\hat{{\bm\beta}})\right)-H_{\hat{{\bm\beta}}}\left(F(y|\mathbf{x},\hat{{\bm\beta}})\right)\right]-{\sqrt{n}}\left[\hat{H}_{{  \bm\beta}_0}\left(F(y|\mathbf{x},{{\bm\beta}_0})\right)-H_{{\bm\beta}_0}\left(F(y|\mathbf{x},{\bm\beta}_0)\right)\right]\\&+{\sqrt{n}}\left[{H}_{\hat{  \bm\beta}}\left(F(y|\mathbf{x},\hat{{\bm\beta}})\right)-H_{{\bm\beta}_0}\left(F(y|\mathbf{x},{\bm\beta}_0)\right)\right]
		\\=&(A)+(B)+(C),
	\end{split}
\end{align}
where \begin{gather*}
	(A)={\sqrt{n}}\left[\hat{H}_{{  \bm\beta}_0}\left(F(y|\mathbf{x},{{\bm\beta}_0})\right)-H_{{\bm\beta}_0}\left(F(y|\mathbf{x},{\bm\beta}_0)\right)\right]\\
	(B)={\sqrt{n}}\left[\hat{H}_{\hat{{\bm\beta}}}\left(F(y|\mathbf{x},\hat{{\bm\beta}})\right)-H_{\hat{{\bm\beta}}}\left(F(y|\mathbf{x},\hat{{\bm\beta}})\right)\right]-{\sqrt{n}}\left[\hat{H}_{{  \bm\beta}_0}\left(F(y|\mathbf{x},{{\bm\beta}_0})\right)-H_{{\bm\beta}_0}\left(F(y|\mathbf{x},{\bm\beta}_0)\right)\right]\\
	(C)={\sqrt{n}}\left[{H}_{\hat{  \bm\beta}}\left(F(y|\mathbf{x},\hat{{\bm\beta}})\right)-H_{{\bm\beta}_0}\left(F(y|\mathbf{x},{\bm\beta}_0)\right)\right].
\end{gather*}
In the preceding display, the first  term $(A)$ corresponds to the behavior of the residuals  when the underlying parameters $\bm\beta_0$ are hypothetically known. We will show that the  second term $(B)$ converges to 0 in probability. The last one  $(C)$ is a drift term induced by the uncertainties in $\hat{{\bm\beta}}.$

To prove  Theorem \ref{theorem}, we first present two lemmas.
\begin{lemma}\label{lemma:b1}
	The class of functions 
	\begin{align*}
		\mathcal{F}=	\left\lbrace 
		\mathbf{x}\rightarrow f_{s,{\bm\beta}}	(\mathbf{x})=s1(p_0(\mathbf{x},{\bm\beta})\leq s),s\in(0,1),{\bm\beta}\in B
		\right\rbrace
	\end{align*}is a Donsker class.
\end{lemma}

 Lemma  \ref{lemma:b1} can be established straightforwardly, as $p_0(\mathbf{x},\bm{\beta})$ is a monotone function of $\mathbf{x}'\bm{\beta}$.

\begin{lemma}\label{lemma:b2}
	By Assumptions \ref{op}, \ref{assume:margin} and \ref{lips}, for $(\mathbf{x},y)$ in the sample space,
	$$
	\mathrm{E}_{\mathbf{X}_i}\left[F(y|\mathbf{x},{\bm\beta_0})1 (p_0(\mathbf{X}_i,{\bm\beta_0})\leq F(y|\mathbf{x},{\bm\beta_0}))-F(y|\mathbf{x},\hat{\bm\beta})1 (p_0(\mathbf{X}_i,\hat{\bm\beta})\leq F(y|\mathbf{x},\hat{\bm\beta}))\right]^2\rightarrow0.$$
\end{lemma}
\begin{proof}
	Since $\left|F(y|\mathbf{x},{\bm\beta}_0)1 (p_0(\mathbf{X}_i,{\bm\beta}_0)\leq F(y|\mathbf{x},{\bm\beta_0}))-F(y|\mathbf{x},\hat{\bm\beta})1 (p_0(\mathbf{X}_i,\hat{\bm\beta})\leq F(y|\mathbf{x},\hat{\bm\beta}))\right|$ is bounded by 1, it suffices to show that 
	$$
	\mathrm{E}_{\mathbf{X}_i}\left|F(y|\mathbf{x},{\bm\beta}_0)1 (p_0(\mathbf{X}_i,{\bm\beta}_0)\leq F(y|\mathbf{x},{\bm\beta_0}))-F(y|\mathbf{x},\hat{\bm\beta})1 (p_0(\mathbf{X}_i,\hat{\bm\beta})\leq F(y|\mathbf{x},\hat{\bm\beta}))\right|\rightarrow0.$$
	We write 
	\begin{align*}
&\mathrm{E}_{\mathbf{X}_i}\left|F(y|\mathbf{x},{\bm\beta}_0)1 (p_0(\mathbf{X}_i,{\bm\beta}_0)\leq F(y|\mathbf{x},{\bm\beta}_0))-F(y|\mathbf{x},\hat{\bm\beta})1 (p_0(\mathbf{X}_i,\hat{\bm\beta})\leq F(y|\mathbf{x},\hat{\bm\beta}))\right|\\
\leq &\mathrm{E}_{\mathbf{X}_i}\left|F(y|\mathbf{x},{\bm\beta}_0)1 (p_0(\mathbf{X}_i,{\bm\beta}_0)\leq F(y|\mathbf{x},{\bm\beta}_0))-F(y|\mathbf{x},{\bm\beta}_0)1 (p_0(\mathbf{X}_i,\hat{\bm\beta})\leq F(y|\mathbf{x},{\bm\beta}_0))\right|\\&+\mathrm{E}_{\mathbf{X}_i}\left|F(y|\mathbf{x},{\bm\beta}_0)1 (p_0(\mathbf{X}_i,\hat{\bm\beta})\leq F(y|\mathbf{x},{\bm\beta}_0))-F(y|\mathbf{x},{\bm\beta}_0)1 (p_0(\mathbf{X}_i,\hat{\bm\beta})\leq F(y|\mathbf{x},\hat{\bm\beta}))\right|
\\&+\mathrm{E}_{\mathbf{X}_i}\left|F(y|\mathbf{x},{\bm\beta}_0)1 (p_0(\mathbf{X}_i,\hat{\bm\beta})\leq F(y|\mathbf{x},\hat{\bm\beta}))-F(y|\mathbf{x},\hat{\bm\beta})1 (p_0(\mathbf{X}_i,\hat{\bm\beta})\leq F(y|\mathbf{x},\hat{\bm\beta}))\right|\\
\leq &\mathrm{E}_{\mathbf{X}_i}\left|1 (p_0(\mathbf{X}_i,{\bm\beta}_0)\leq F(y|\mathbf{x},{\bm\beta}_0))-1 (p_0(\mathbf{X}_i,\hat{\bm\beta})\leq F(y|\mathbf{x},{\bm\beta}_0))\right||\\&+\mathrm{E}_{\mathbf{X}_i}\left|1 (p_0(\mathbf{X}_i,\hat{\bm\beta})\leq F(y|\mathbf{x},{\bm\beta}_0))-1 (p_0(\mathbf{X}_i,\hat{\bm\beta})\leq F(y|\mathbf{x},\hat{\bm\beta}))\right|\\&+\mathrm{E}_{\mathbf{X}_i}\left|F(y|\mathbf{x},{\bm\beta}_0)-F(y|\mathbf{x},\hat{\bm\beta})\right|.
	\end{align*}
The last term is bounded by $|\hat{{\bm\beta}}-\bm{\beta}_0|$ by Assumption \ref{lips}.
Now we analyze the first term, which is bounded by 
\begin{align*}
	\mathrm{E}_{\mathbf{X}_i}1\left(\left|p_0(\mathbf{X}_i,{\bm\beta}_0)-F(y|\mathbf{x},{\bm\beta}_0)\right|\leq \left|p_0(\mathbf{X}_i,{\bm\beta}_0)-p_0(\mathbf{X}_i,\hat{\bm\beta})\right|\right).
\end{align*}
By Assumption \ref{lips}, $\left|p_0(\mathbf{X}_i,{\bm\beta}_0)-p_0(\mathbf{X}_i,\hat{\bm\beta})\right|$ is bounded by 
 $|\hat{{\bm\beta}}-\bm{\beta}_0|$ multiplied by a constant.
Furthermore, the density of $p_0(\mathbf{X}_i,{\bm\beta}_0)$ is bounded by Assumption  \ref{assume:margin}. Hence, the first term is of order  $|\hat{{\bm\beta}}-\bm{\beta}_0|$.

The second term can be handled in a similar manner and bounded by
\begin{align*}
	\mathrm{E}_{\mathbf{X}_i}1\left(\left|p_0(\mathbf{X}_i,\hat{\bm\beta})-F(y|\mathbf{x},\hat{\bm\beta})\right|\leq \left|F(y|\mathbf{x},\hat{\bm\beta})-F(y|\mathbf{x},{\bm\beta}_0)\right|\right).
\end{align*}
By Assumptions  \ref{assume:margin} and \ref{lips}, this  is also bounded by $|\hat{{\bm\beta}}-\bm{\beta}_0|$ multiplied by a constant.
The  lemma follows upon combining the above results and 
 Assumption \ref{op}.
\end{proof}

\begin{proof}[Proof of Theorem \ref{theorem}]
	Define $$X_n(s,{\bm\beta})=\frac{1}{\sqrt{n}}\sum_{i=1}^{n}\left[s1(p_0(\mathbf{X}_i,{\bm\beta})\leq s)-s\Pr (p_0(\mathbf{X}_i,{\bm\beta})\leq s)\right],$$which weakly converges to a Gaussian process 
	$\mathbb{G}
	f_{s,\bm{\beta}}$
	from  Lemma \ref{lemma:b1}.
For $(\mathbf{x},y)$ in the sample space, define
	a mapping from $l^\infty((0,1)\times B)\rightarrow l^\infty( B)$ $${X_n}(s,{\bm\beta})\rightarrow Z_n(\mathbf{x},y,{\bm\beta})\coloneqq X_n(F(y|\mathbf{x},{\bm\beta}),\bm\beta).$$
Since $$\big\|Z_n(\mathbf{x},y,\cdot)-Z_n'(\mathbf{x},y,\cdot)\big\|_{ B}=\sup_{\bm{\beta}\in B}\left|X_n(F(y|\mathbf{x},{\bm\beta}),{\bm\beta})-X_n'(F(y|\mathbf{x},{\bm\beta}),{\bm\beta})\right\|\leq ||X_n-X_n'||_{(0,1)\times B},$$ this  is a continuous mapping.
	By the continuous mapping theorem,  
	$Z_n(\mathbf{x},y,\cdot)$ converges weakly in $l^\infty( B)$. 
	Consequently, $(A)$ in  \eqref{eq:hn} converges to $\mathbb{G}
	f_{F(y|\mathbf{x},{\bm\beta}_0),\bm{\beta}_0},$ where $f_{s,\bm{\beta}}$ is defined in Lemma \ref{lemma:b1}.

	By asymptotic tightness, $Z_n(\mathbf{x},y,\cdot)$
	has continuous sample paths with respect to the semimetric $\rho$ given by
	\begin{align*}
	\rho(\bm{\beta},\bm{\beta}')
	=&\mathrm{E}_{\mathbf{X}_i}	\left[f_{F(y|\mathbf{x},{\bm\beta}),\bm{\beta}}(\mathbf{X}_i)-f_{F(y|\mathbf{x},{\bm\beta}'),\bm{\beta}'}(\mathbf{X}_i)\right]^2\\
	=&\mathrm{E}_{\mathbf{X}_i}\left[F(y|\mathbf{x},{\bm\beta})1 (p_0(\mathbf{X}_i,{\bm\beta})\leq F(y|\mathbf{x},{\bm\beta}))-F(y|\mathbf{x},{\bm\beta}')1 (p_0(\mathbf{X}_i,{\bm\beta}')\leq F(y|\mathbf{x},{\bm\beta}'))\right]^2
	.
\end{align*}
	 By Lemma \ref{lemma:b2}, $\rho(\bm{\beta}_0,\hat{\bm{\beta}})\rightarrow0.$ Therefore,
	$\left|Z_n(\mathbf{x},y,{\bm\beta}_0)-Z_n(\mathbf{x},y,\hat{{\bm\beta}})\right|$, which is the second term  $(B)$ of \eqref{eq:hn}, converges to 0 in probability.
	
	Now we show the convergence of $(C)$ in \eqref{eq:hn},
	$\sqrt{n}\left(H_{\hat{  \bm\beta}}\left(F(y|\mathbf{x},\hat{{\bm\beta}})\right)-H_{{\bm\beta}_0} \left( F(y|\mathbf{x},{\bm\beta}_0)\right)\right).$
	By Assumption \ref{op},    $$\sqrt{n}\left(\hat{  \bm\beta}-\bm\beta_{0}\right)=\left[I(\bm\beta_0)\right]^{-1}\frac{1}{\sqrt{n}}\sum_{i=1}^n\dot{l}_{\bm\beta_{0}}(\mathbf{X}_i,Y_{i})+o_p(1).$$ 
	That is, $\hat{ \bm \beta}$ is asymptotically linear.
	Furthermore, $P\dot{l}_{\bm\beta_0}=0$. The functional delta method together with	Assumption \ref{monof} gives 
	\begin{align*}
&\sqrt{n}\left(H_{\hat{  \bm\beta}}\left(F(y|\mathbf{x},\hat{{\bm\beta}})\right)-H_{{\bm\beta}_0} \left( F(y|\mathbf{x},{\bm\beta}_0)\right)\right)\\=&
	\sqrt{n}\left(\hat{ \bm \beta}-\bm\beta_{0}\right)\left.\frac{\partial  H_{{\bm\beta}} \left( F(y|\mathbf{x},{\bm\beta})\right)}{\partial \bm\beta}\right\vert_{\bm\beta=\bm\beta_0}\\=&\left.\frac{\partial H_{{\bm\beta}} \left( F(y|\mathbf{x},{\bm\beta})\right)}{\partial \bm\beta}\right\vert_{\bm\beta=\bm\beta_0}\left[I(\bm\beta_0)\right]^{-1}\mathbb{G}_n\dot{l}_{\bm\beta_{0}}+o_p(1).\end{align*}

Finally, the joint limit law of  (A) and (C) can be determined from the marginals, and the limit of the sum of the two terms can be represented in the form as given in Theorem \ref{theorem}.
\end{proof}

\end{document}